\documentclass[12pt]{elsart}

\usepackage{float}
\usepackage{epsfig}
\usepackage{amsfonts}
\usepackage{amssymb}

\begin{document}

\begin{frontmatter}
\title{Evolution and anti-evolution in a minimal stock market model}
\author{R. Rothenstein\thanksref{CORR}},
\author{K. Pawelzik}
\address{Institut f\"ur Theoretische Physik, Universit\"at Bremen, Otto-Hahn-Allee 1, D-28334 Bremen, Germany}
\thanks[CORR]{Corresponding author. Institut f\"ur Theoretische Physik, Universit\"at Bremen, Otto-Hahn-Allee 1, D-28334 Bremen, Germany. Tel.:(+49)-421-218-4526; Fax: (+49)-421-218-9104. Email address: rroth@physik.uni-bremen.de} 

\begin{abstract}
We present a novel microscopic stock market model consisting of a
large number of random agents modeling traders in a market. Each
agent is characterized by a set of parameters that serve to make
iterated predictions of two successive returns. The future price 
is determined according to the offer and the demand of all agents. 
The system evolves by redistributing the capital among the
agents in each trading cycle.
Without noise the dynamics of this system is nearly regular and thereby
{\em fails} to reproduce the stochastic return fluctuations observed in
real markets. However, when in each cycle a small amount of
noise is introduced we find the typical features of real financial time 
series like fat-tails of the return distribution and large temporal 
correlations in the volatility without significant correlations in the 
price returns.
Introducing the noise by an evolutionary process leads to different 
scalings of the return distributions that depend on the definition of
fitness. 
Because our realistic model has only very few parameters, and the
results appear to be robust with respect to the noise level and the 
number of agents we expect that our framework may serve as new paradigm 
for modeling self generated return fluctuations in markets.
\end{abstract}

%\pacs{01.75.+m; 02.50.Le; 05.40.-a; 87.23.Ge}
\begin{keyword}
Econophysics; Agent-based model; Stock market model; Evolution; Fat tails 
\end{keyword}
\end{frontmatter}

\section{Introduction}

Empirical studies of stock markets and foreign exchange rates
demonstrate that financial time series exhibit some universal
characteristics
\cite{Mandelbrot_63,Mantegna_95,Ghashghaie_96,Bouchaud,Mantegna}.
The distribution of returns usually has fat-tails and there are 
large temporal correlations in the volatility (``volatility 
clusters'') without correlation in the returns. These findings 
are in contrast to the hypothesis that, due to independence of 
market traders financial time series are simple random walks 
\cite{Bachelier}. This raises the question which underlying 
dynamics is responsible for these properties.

Microscopic models developed in recent years
\cite{Caldarelli_97,Lux_99,Levy_95,Levy,Palmer_94,Maslov_00}
attempted to reproduce the essential features of real stock market
time series. Although these models reproduced some statistical 
aspects of price time series, a deeper understanding of the 
underlying processes still remains difficult, mainly, because most 
of the models are not very robust with respect to the choice of 
parameters and some even depend on the number of agents in the 
system \cite{Egenter_99}.

Starting from a different motivation, game theoretical models
\cite{Arthur_94,Challet_97,Donangelo_00,Cont_00} were considered to
understand the interplay between different players in an artifical
market. These models have only a few parameters and allow an analytical
understanding of interaction in multi-agent-models. The simplicity  
of these models, however makes an application to real markets difficult.

In this paper we present a simple paradigmatic model which combines
 these two approaches. In particular we apply a game theoretical 
Ansatz with only a very small number of free parameters and a price 
determination close to reality \cite{Busshaus_99}. This is a necessary 
step towards understanding the mechanisms that lead to fat-tails and 
volatility clusters and may enable the analysis of the dynamics. In this 
paper we concentrate on the effect that different evolution strategies 
have on the statistical properties of price fluctuations.

In section 2 we describe the basic model. In section 3 we show that
our model naturally reproduces the fat-tails of return distributions
and the emergence of volatility clusters.
In section 4 we introduce evolution as a source of noise
into the model and finally show how different selection mechanisms
influence the shape of the return distributions.

\section{Basic model}

In our model of a stock market $N$ agents trade by swapping stocks
into cash and vice versa. Initially every agent $i$ receives a
number of $S_0$ stocks and an amount $M_0$ of cash. The decision of
one agent to buy or to sell stocks is determined by a set of P
parameters $\alpha_{\Delta t}^i (\Delta t=0,...,\tau)$. These
parameters define the linear prediction model of each agent and are
randomly drawn from a normal distribution with mean 0 and variance
$\sigma^2$ ($\mathcal{N}(0,\sigma)$).

Every simulation cycle of the model consists of three steps: In the
first step, the agents make a prognosis of two successive future 
returns based on their individual prediction model. In the second 
step the overall demand and supply of the stocks is calculated and 
the new price is determined according to an order book. In the third 
step all possible orders are executed and the strategies of all agents 
are rearranged by a perturbation of their parameters.

\subsection{Prediction of the future returns}

For the prediction of the future price each agent only takes the
log-returns $\ln(\frac{p(t-1)}{p(t-2)})=r(t-1)$ of the price
history p(t) into account, which ensures that the absolute value of
a stock price is not relevant.

For simplicity we assume that every agent makes only a linear
prediction of the two following future returns by weighting the
past $P= \tau + 1$ returns of the time series with her individual
parameters $\alpha$. At time $t-1$ an agent predicts the returns
$\hat{r}(t)$ and $\hat{r}(t+1)$ using the following equations:
\begin{eqnarray*}
    \hat{r}^i_t = f^i(r_{t-1},r_{t-2},...,r_{t-\tau}) =
    \alpha^i_0 +\sum^{\tau}_{\Delta t=1} \alpha^i_{\Delta t} r(t-\Delta t)
\end{eqnarray*}
and
\begin{eqnarray*}
    \hat{r}^i_{t+1} = f^i(\hat{r}^i_t,r_{t-1},...,r_{t -\tau+1})=
    \alpha^i_0 + \alpha^i_1 \hat{r}^i_t
    + \sum^{\tau}_{\Delta t=2} \alpha^i_{\Delta t} r(t-\Delta t+1)
\end{eqnarray*}

For both successive predictions the agents use the same
deterministic linear prediction function $f^i$, clearly using for the 
second prediction the outcome of the first, i.e. the strategy of an agent
$i$ is fully determined by the $P$ prediction parameters
$(\alpha^i_0,\alpha^i_1,\alpha^i_2,...\alpha^i_{\tau})$.

\subsection{Determination of the new price}
\label{sec:Determination of the new price}

After all agents made their predictions the decision whether an
agent sells or buys shares is made: If $ \hat{r}^i_{t+1} <0$ the
agent $i$ makes an offer at the price $\hat{p}^i_t= p_{t-1}
e^{\hat{r}^i_t}$ and becomes a potential seller for this trading 
cycle. Otherwise, i.e. if $ \hat{r}^i_{t+1} >0$, the agent $i$ 
makes a bid at price $\hat{p}^i_t= p_{t-1} e^{\hat{r}^i_t}$ and 
intends to become a buyer. 

This means that if e.g. the agent believes that
the price will go down from time $t$ to time $t+1$ he will try to
sell his stocks, as long as the price at time $t$ is higher than
$\hat{p}^i_t$.

Now two functions are computed to fix the new price: an offer
function with all offers and a demand function containing all bids
(see also \cite{Busshaus_99}).
\begin{eqnarray*}
    O(p)=\sum_{i: \hat{r}^i_{t+1} <0} S^i \Theta(p-\hat{p}^i_t)
\end{eqnarray*}
\begin{eqnarray*}
    D(p)=\sum_{i:\hat{r}^i_{t+1} > 0} \Delta S^i \Theta(\hat{p}^i_t-p)
\end{eqnarray*}

\begin{figure}[ht]
\begin{center}
\includegraphics[width=9cm,angle=270]{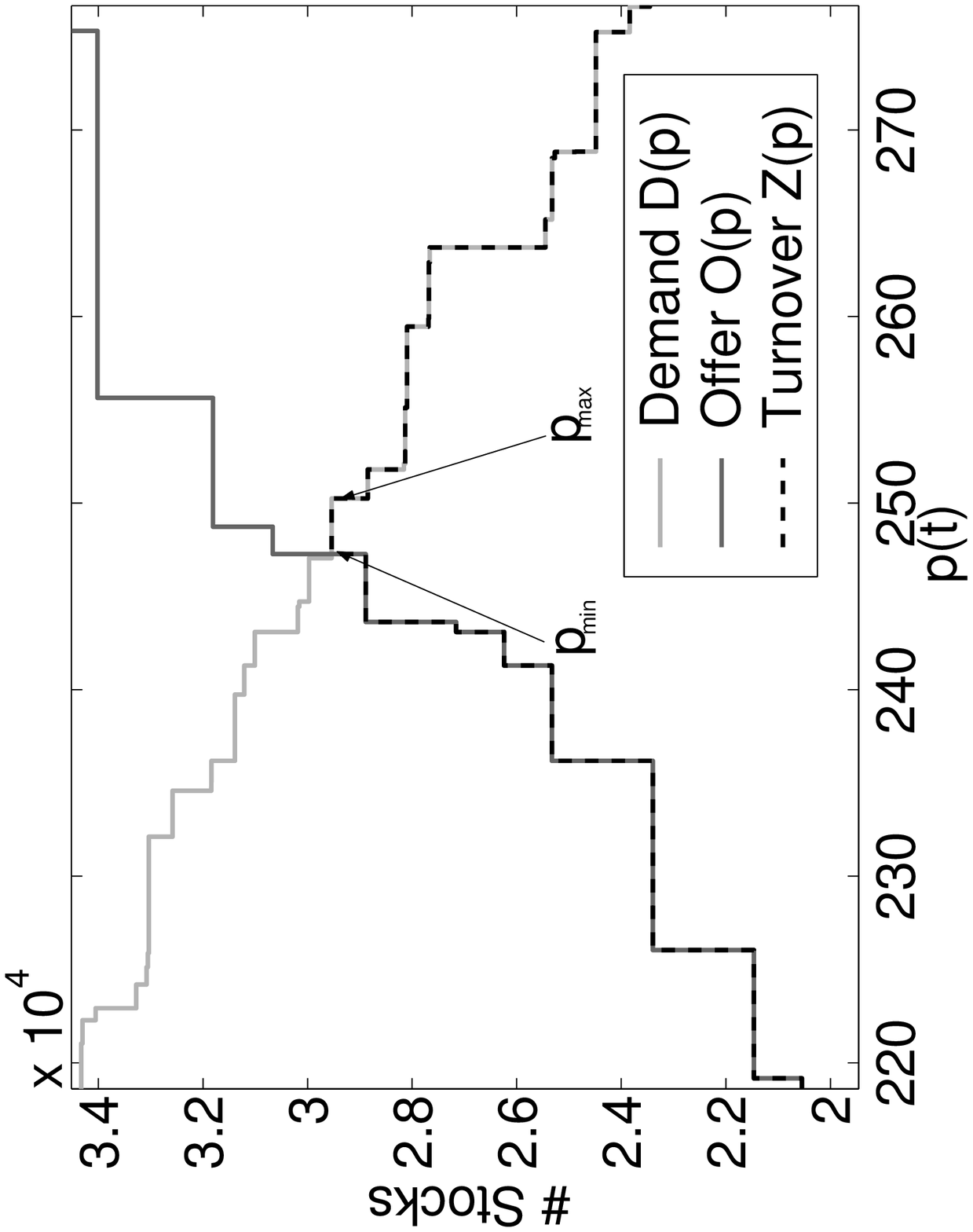} \hfill
\caption{\small The demand D(p), offer O(p) and turnover 
$Z(p)= \min\{D(p),O(p)\}$ as function of the price p in one trading cycle. 
The turnover reflects how many shares would be traded at a certain price. 
The new price is determined in the interval between $p_{min}$ and 
$p_{max}$ where the function Z(p) has its maximum. }
\label{Abb:DemandOffer}
\end{center}
\end{figure}

$S^i$ is the total number of stocks agent $i$ owns. $\Delta S^i=$
int$[\frac{M^i}{\hat{p}^i_t}]$, is the integer number of stocks,
which agent $i$ is able to buy with her money $M^i$. $\Theta(x)$ is
the Heavy-side function with  $\Theta(x)=1$ for $x \ge 0$ and
$\Theta(x)=0$ for $x < 0$. $O(p)$ is the offer function, where all
offers ($ \hat{r}^i_{t+1} <0$) are collected. The function is
monotonously rising with steps at the limit prices, where an agent
$i$ is willing to sell his stocks (if he had at least one). D(p) 
is the respective monotonously decreasing demand function (see
Fig.(\ref{Abb:DemandOffer})) As seen from the equation above every
agent either intends to sell all her stocks or to use all her money
which means that the agents fully believe in their prognosis. Both
functions together represent the order book of our stock market
model.
Now we calculate the minimum of both functions at some price $p$:
\begin{eqnarray*}
Z(p)=\min\{O(p),D(p)\}
\end{eqnarray*}
This functions reflects the transaction volume in stocks at a
certain price $p$, i.e. it represents the turnover function.\\ 
In order to determine the new price we take the minimum and the
maximum argument of $Z(p)$ at the interval of the maximum turnover:
$p_{min} = \min( $argmax $ Z(p))$ and $p_{max} = \max( $argmax $
Z(p))$. The new price is then defined by the weighted mean between
these two points:
$p(t)=\frac{p_{\min}O(p_{\min})+p_{\max}D(p_{\max})}{O(p_{\min})+
D(p_{\max})}$.
Sometimes it happens that their is a spread between the highest bid 
and the lowest offer, therefore the turnover function equals zero for 
all values of $p$. Due to the fact that the standard process of price 
determination lead to no meaningful price in these situation
we alternatively choose the price in the middle between the highest 
bid and the lowest offer.

\subsection{Execution of orders and agent dynamics}
\label{sec:Execution of orders and agent dynamics}

When the new price is fixed, the agents execute their orders. All
buyers with $\hat{p} < p_{min}$ buy $\Delta S^i$ stocks and all
sellers with $\hat{p} > p_{max}$ sell all their stocks $S^i$. At
the point of intersection the offer in general does not match the
bid and therefore here only the difference between offer and bid
can be traded.\\

Before the next cycle is started we add a small amount of noise of
amplitude $\tilde{\sigma}$ to all parameters of the agents. 
With $\xi$ drawn from a normal distribution ($\mathcal{N}(0,1)$) 
the parameters of each agent then are perturbed to yield:
\begin{eqnarray*}
 \alpha_{\Delta t}^i(t+1)=\frac{\sigma}{\sqrt{\sigma^2+\tilde\sigma^2}} 
(\alpha_{\Delta t}^i + \tilde\sigma \xi)
\end{eqnarray*}
The factor $\frac{\sigma}{\sqrt{\sigma^2+\tilde\sigma^2}}$ has been
introduced to keep the variance of the parameter distribution
independent of the choice of $\tilde\sigma$ at the constant value
$\sigma^2$. This ensures that simulations with different
$\tilde\sigma$ can be compared.

\section{Results of the basic model}
\begin{figure}[ht]
\begin{center}
\includegraphics[width=9cm,angle=270]{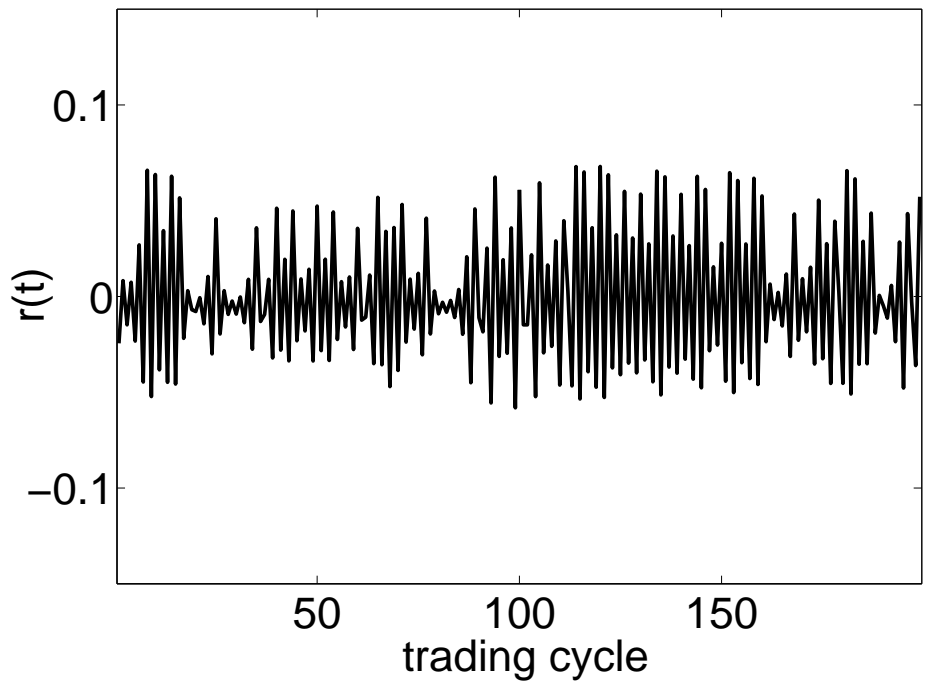}\hfill
\caption{\small Log-returns $r(t)= log(\frac{p(t)}{p(t-1)})$ of a simulation produced 
by the Basic Model with parameters $N=1000, P=3, S(0)=1000,M(0)=100000, \sigma=1, 
\tilde{\sigma}=0$.}
\label{Abb:Basic00}
\end{center}
\end{figure}

The results of the basic model are essentially influenced only 
by the noise added to the parameters in each cycle.   
In order to understand the role of this noise we examined the 
effect of it's amplitude $\tilde{\sigma}$. For $\tilde{\sigma}=0$, 
all agents keep their prediction parameters fixed all the time 
leading to a quite regular time series (Fig.\ref{Abb:Basic00}),  
which reflects the deterministic dynamics for the price determination.
For most of the randomly chosen initial conditions, the
average values of $r(t)$ are close to zero, while occasionally it is
 larger than the amplitude of fluctuations.

\begin{figure}[ht]
%\begin{center}
\includegraphics[width=5.5cm,angle=270]{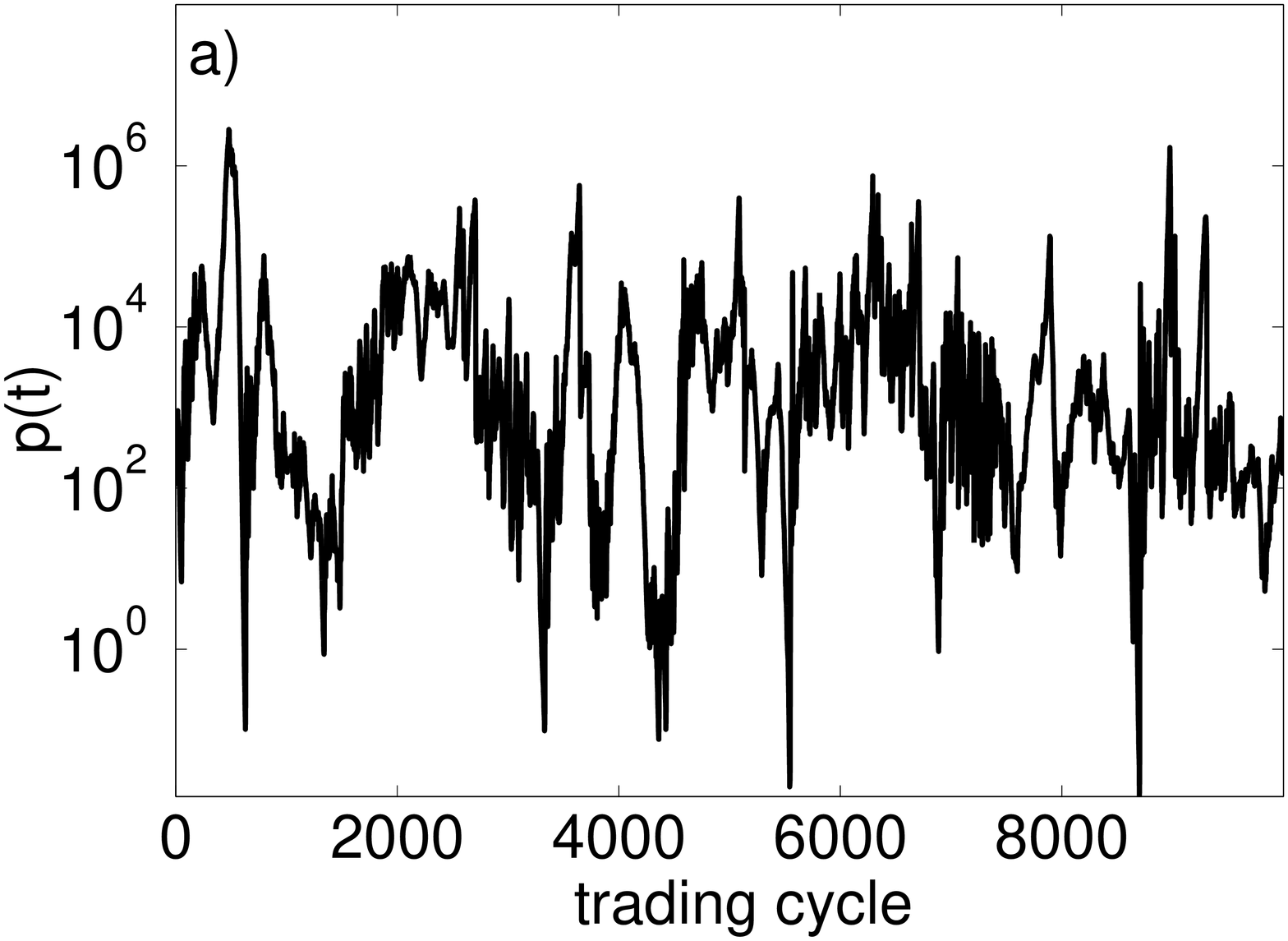} \hfill
\includegraphics[width=5.5cm,angle=270]{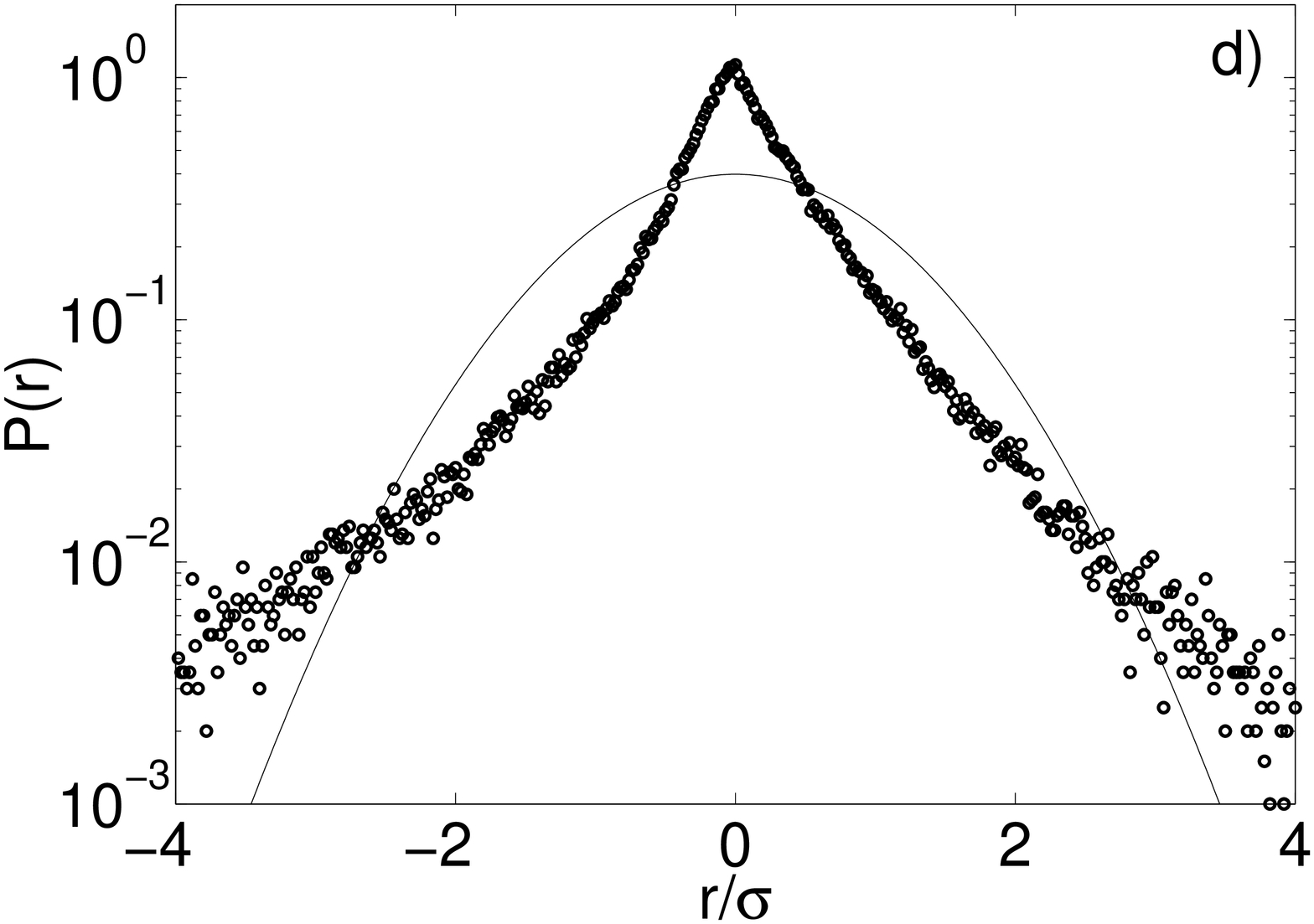} \hfill
\includegraphics[width=5.5cm,angle=270]{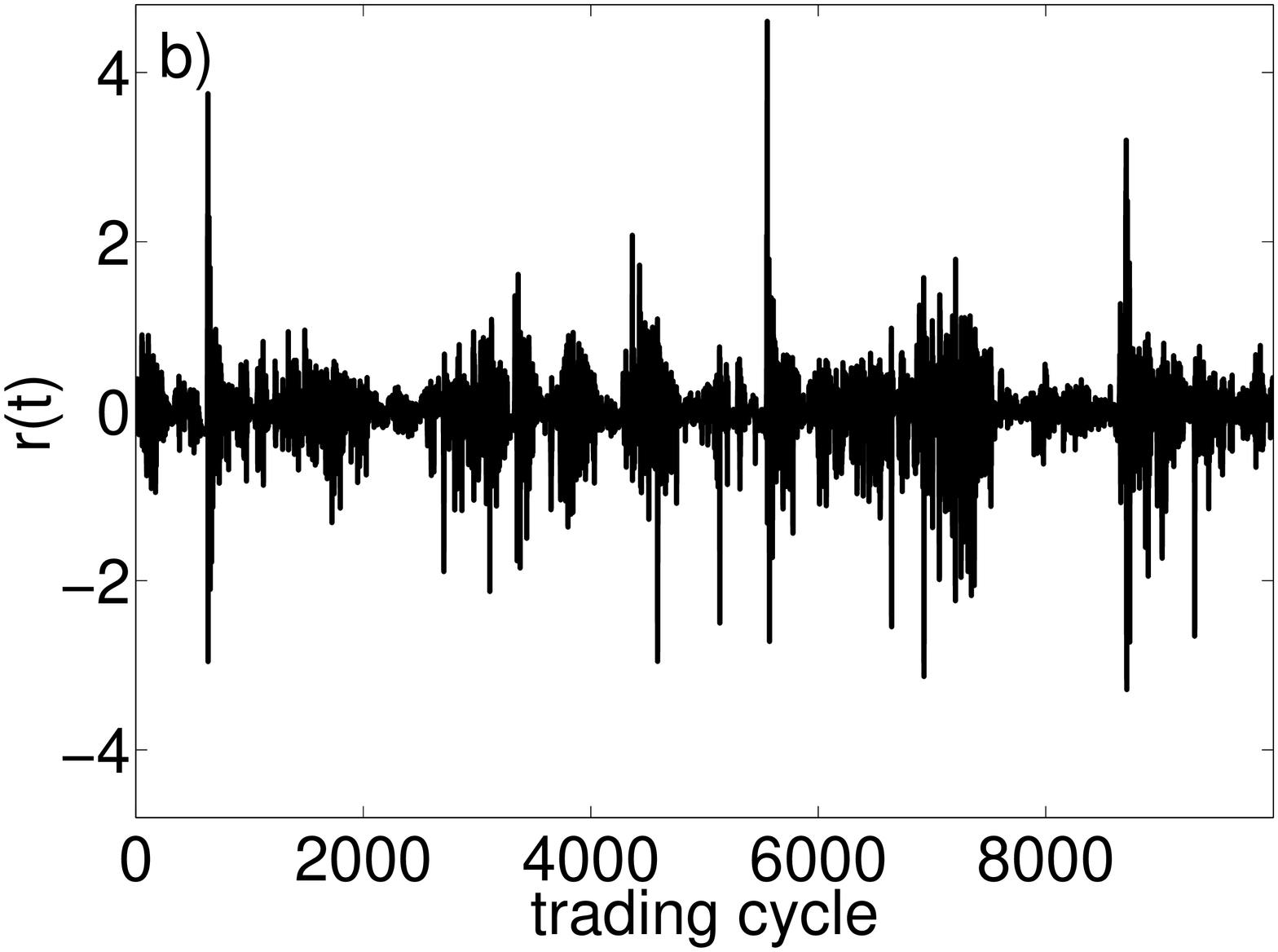} \hfill
\includegraphics[width=5.5cm,angle=270]{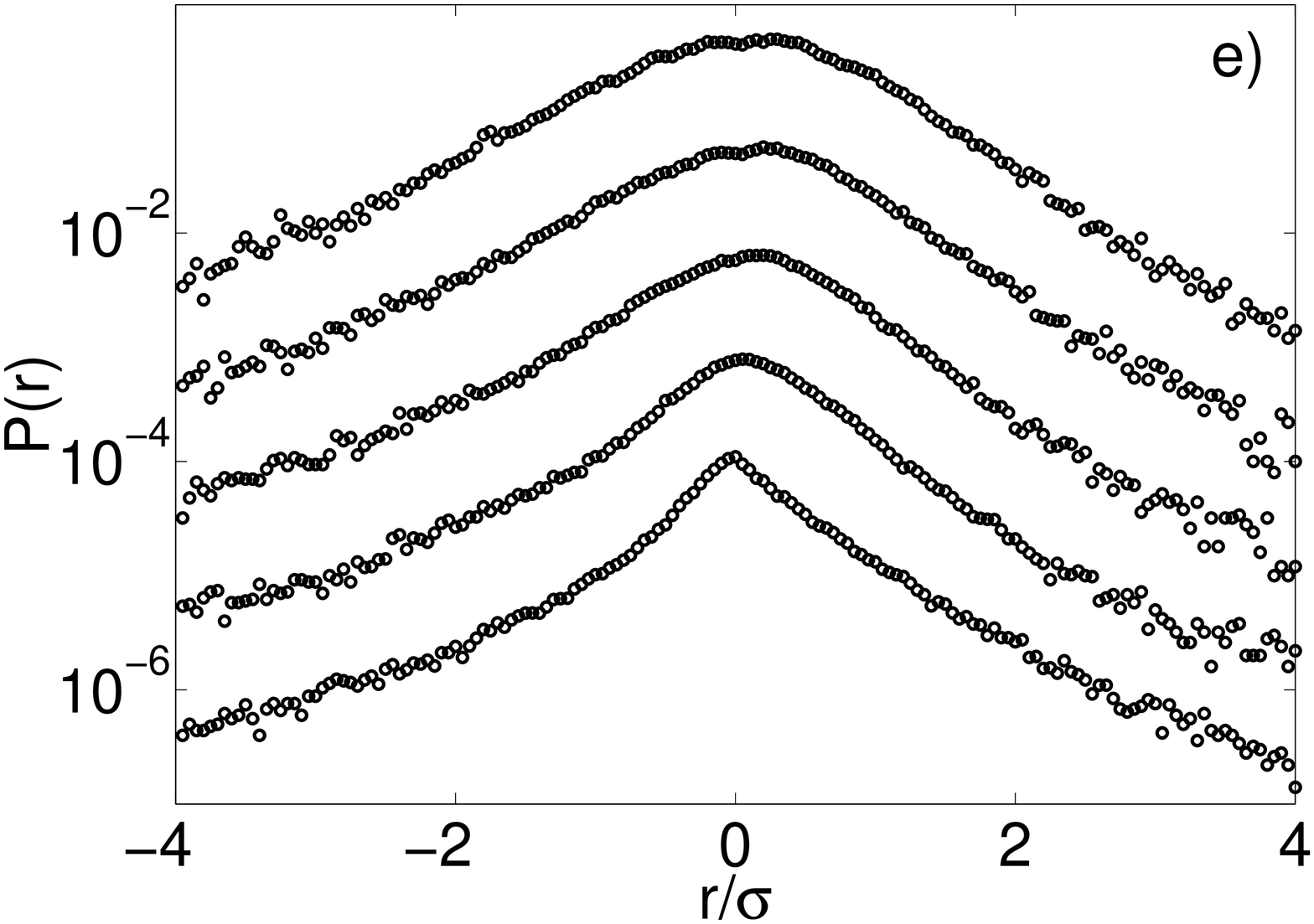} \hfill
\includegraphics[width=5.5cm,angle=270]{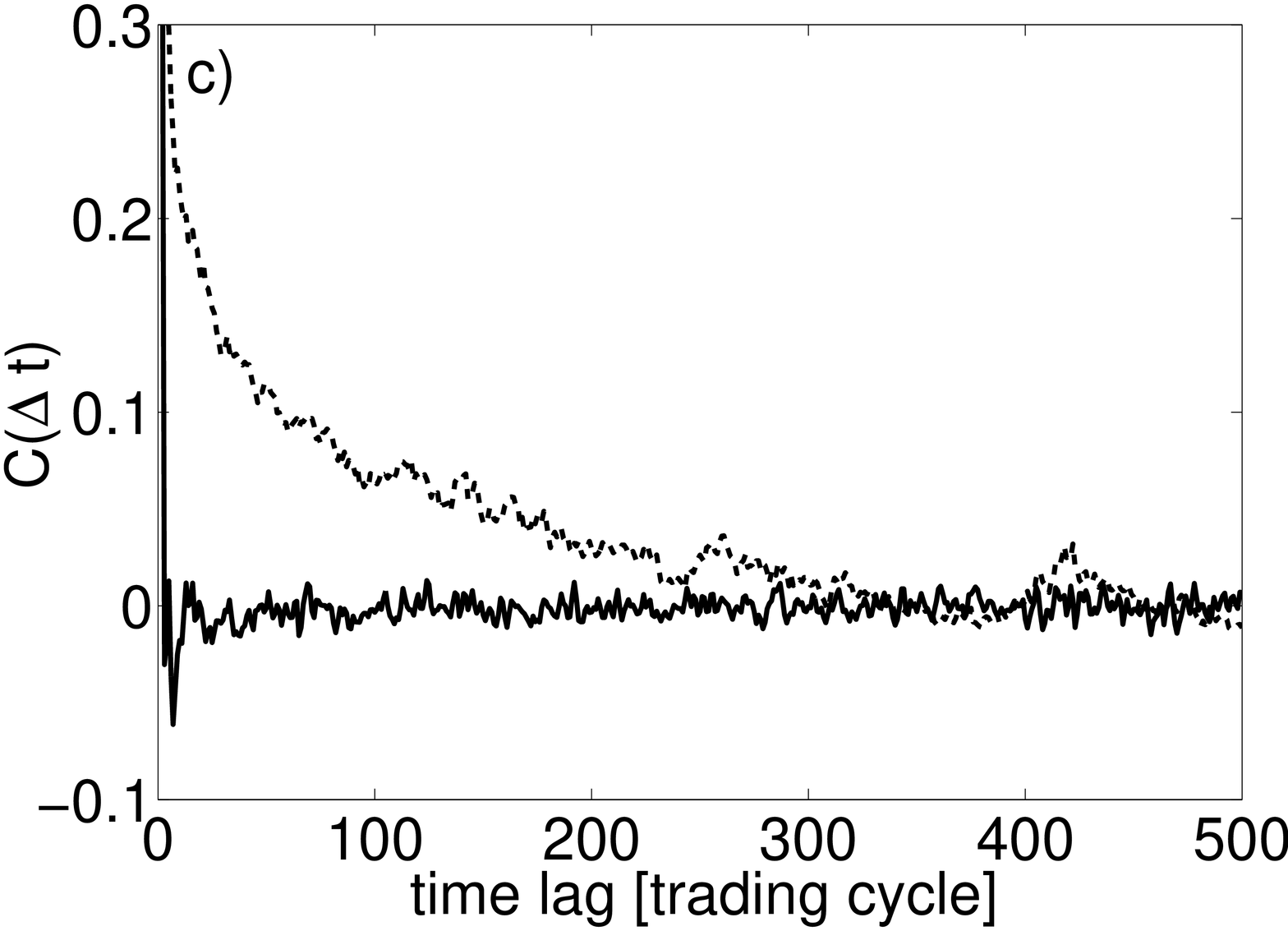} \hfill
\includegraphics[width=5.5cm,angle=270]{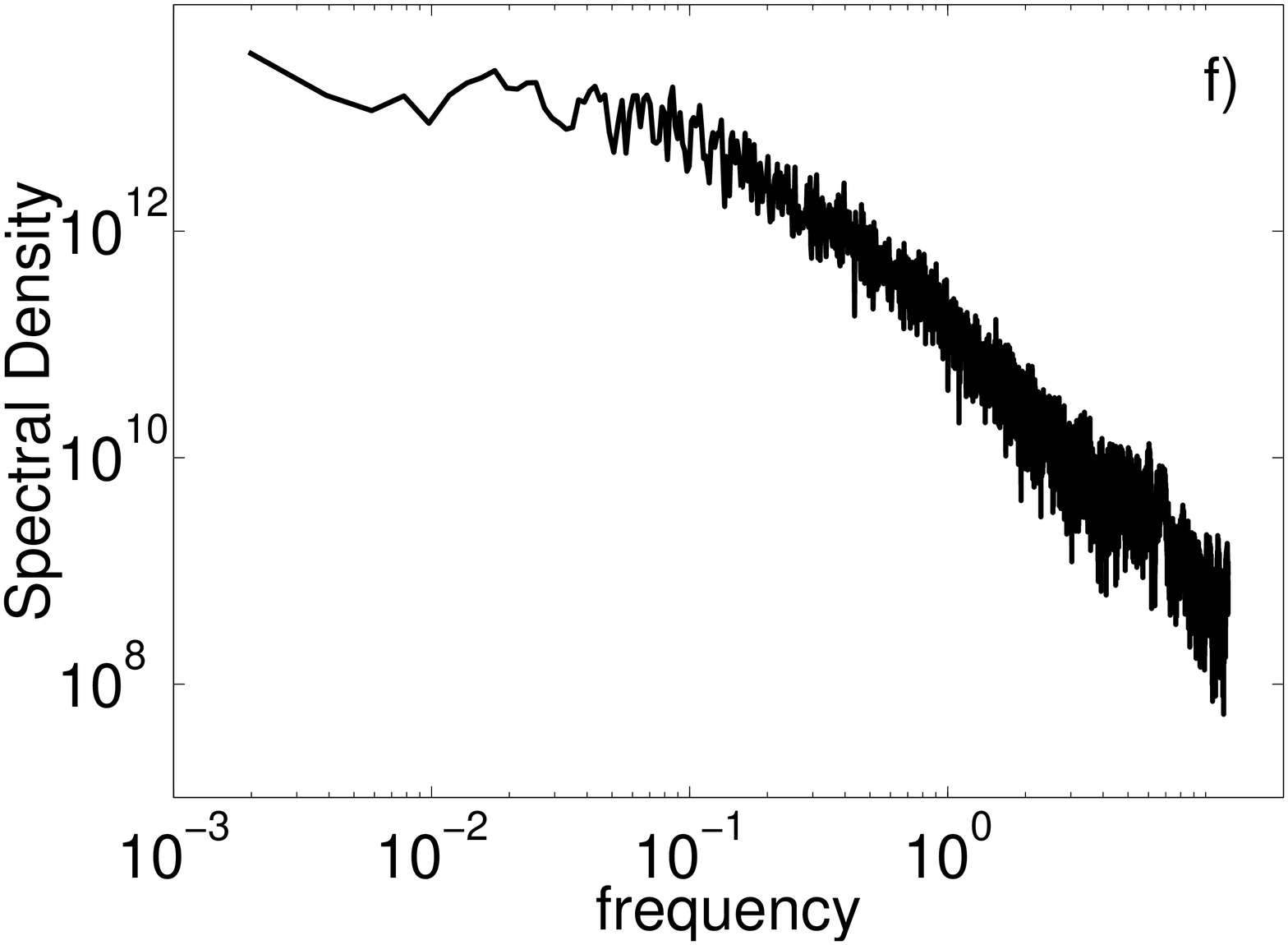} \hfill

\caption{\small Simulation results obtained from the Basic Model with 
parameters $N=1000, P=3, S(0)=1000, M(0)=100000, \sigma=1, \tilde{\sigma}=0.01$.
A minimum of 50000 trading cycles have been discarded in order to avoid effects 
due to transients.
 a) Price history $p(t)$ over 10000 trading cycles.
 b) Log-returns $r(t)= log(\frac{p(t)}{p(t-1)})$ with respect to the price history 
shown in a).
 c) Auto-covariance $C_x(\Delta t)=\frac{\sum_t (x(t)-<x(t)>)
    (x(t+\Delta t)-<x(t+\Delta t)>)}{\sum_t (x(t)-<x(t)>)^2}$ 
    determined on the basis of 50000 trading cycles. 
    The dashed line shows the auto covariance 
    for the absolute values of log-returns $x=|r(t)|$, the solid line shows 
    the auto covariance for $x=r(t)$.
 d) Normalized distribution of log-returns $P(r)$ for $\Delta t =1$ 
    on a semilog plot. For comparison, the solid line shows the 
    standard normal distribution.
 e) Normalized distribution of log-returns $P(r)$ for $\Delta t =1,4,9,16,25$.
    Curves are shifted in vertical direction for the sake of clarity.
 f) Powerspectra (mean over 4 intervals of 10000 trading cycles). 
}
\label{Abb:Basic01}
%\end{center}
\end{figure}

While in the complete absence of noise the model behaves incompatible to
the dynamics of real stock markets, a small amount of noise is sufficient
to reproduces typical features of real stock markets (Fig.\ref{Abb:Basic01}).
 The simulation clearly exhibits the phenomenon 
of volatility clustering (Fig.\ref{Abb:Basic01}b) and therefore induces 
correlations in the absolute log-returns of the time series 
(dashed line in Fig.\ref{Abb:Basic01}c). The correlations of the 
raw returns show only a small anti-correlation 
(solid line in Fig.\ref{Abb:Basic01}c), which is far too
small to be responsible for the correlations in the absolute
returns. Furthermore we see a non-Gaussian shape of the return
distribution (Fig.\ref{Abb:Basic01}d) with fat tails that 
vanish at longer time scales (Fig.\ref{Abb:Basic01}e).
The power spectrum (Fig.\ref{Abb:Basic01}f) of the price exhibits scaling 
with an exponent slightly below two, implying that the Hurst exponent 
$ H \lesssim \frac{1}{2}$.

For larger $\tilde{\sigma}$ the parameters of the agents are perturbed 
stronger during each time step which leads to a smooth change between 
fat-tailed and Gaussian distributions (Fig. \ref{Abb:Multisigma}). 
This shows that the memory in the parameters and the capital of the agents
is a necessary requirement for fat-tails in our model. If the noise on 
these parameters is too strong the trading behavior of the agents becomes 
random and we finally observe the expected Gaussian distribution for 
$\tilde \sigma$. We found that the noise level sufficient to yield the 
behavior described above depends on the number of agents - the more agents 
the less noise keeps the system from becoming regular. 
Also, the power law behavior breaks down if $P \le 2$ - for all simulations
with $P \ge 3$ we saw complex dynamics similar to the results shown above.  

\section{Evolution as noise source}

Here we discuss a simple variant of our model, which can be seen in
analogy to an evolutionary process. Instead of perturbing the
parameters of all agents as described in section \ref{sec:Execution
of orders and agent dynamics} we in each cycle choose one agent
which we eliminate and introduce a new agent with new random
parameters into the market. All other parts of the basic model
remain unchanged.

We consider three different strategies: 1. Eliminate the poorest,
2. the richest or 3. a random one. 1. and 2. are based on the
capital an agent owns $(C^i(t)=M^i(t)+p(t)S^i(t))$ and represent an
'evolutionary' and an 'anti-evolutionary' mechanism of selection.
The new agent starts with money $M_0$ and shares $S_0$ and receives
new prediction parameters $\alpha^i$ randomly drawn from a Gaussian
distribution ($\mathcal{N}(0,\sigma)$). The noise introduced by
this evolutionary mechanism is different from the basic model in
two respects: Firstly, in a model with a larger number of
agents this procedure corresponds to a smaller relative noise 
level in the system. Second, we now not only have noise in the 
parameter space, but instead also introduce noise in the money and the stocks.

At the end of each trading period the amount of stocks and cash are
both normalized by $\sum_{i=1}^N M^i =N M_0$ and $\sum_{i=1}^N S^i= 
N S_0$ to remain comparable to the basic model. This could be
interpreted as some sort of tax that each agent has to pay for letting
the new agent enter the market.

\section{Results of the evolutionary models}

\begin{figure}[ht]
%\begin{center}
\includegraphics[width=5.5cm,angle=270]{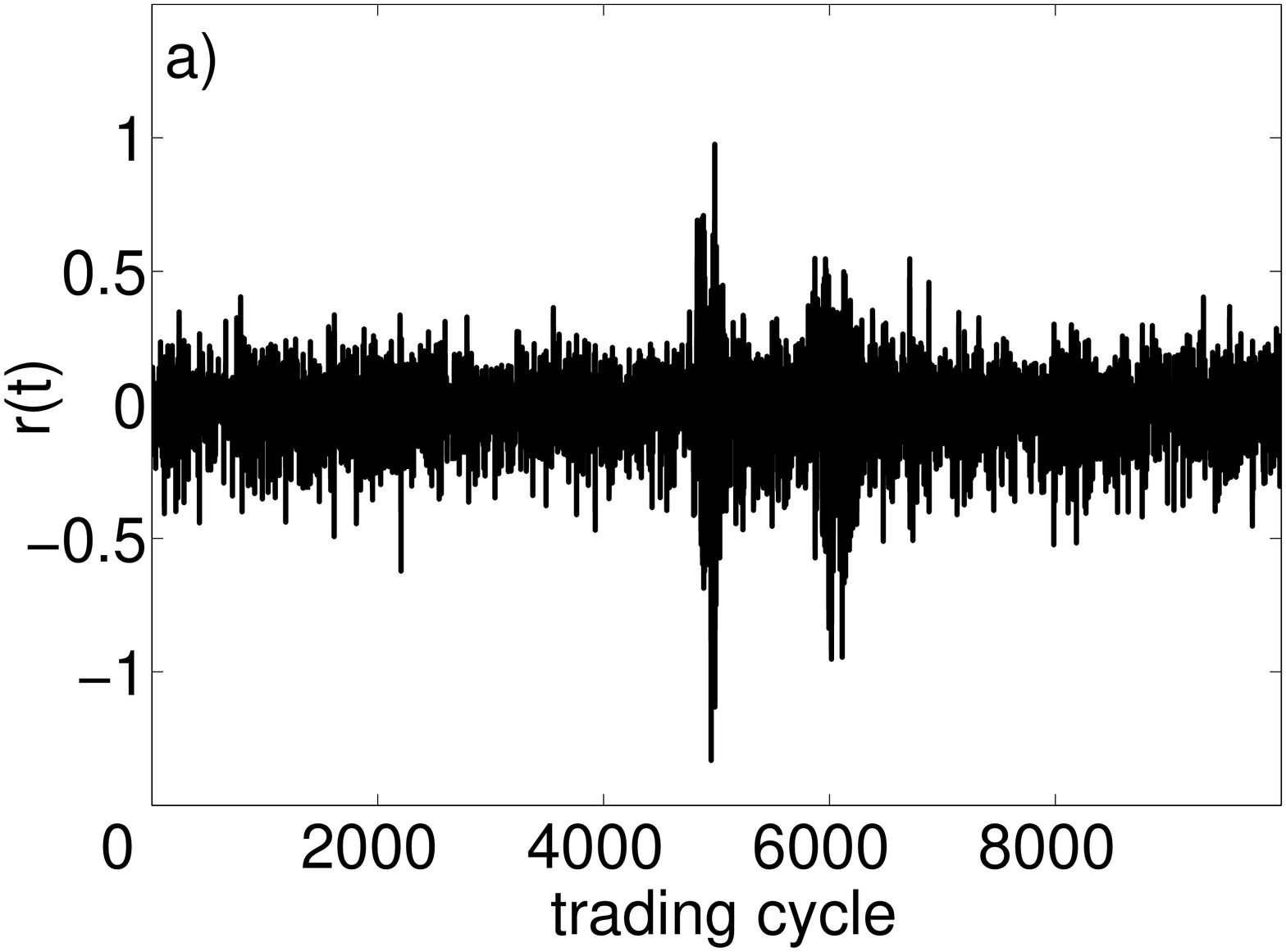} \hfill
\includegraphics[width=5.5cm,angle=270]{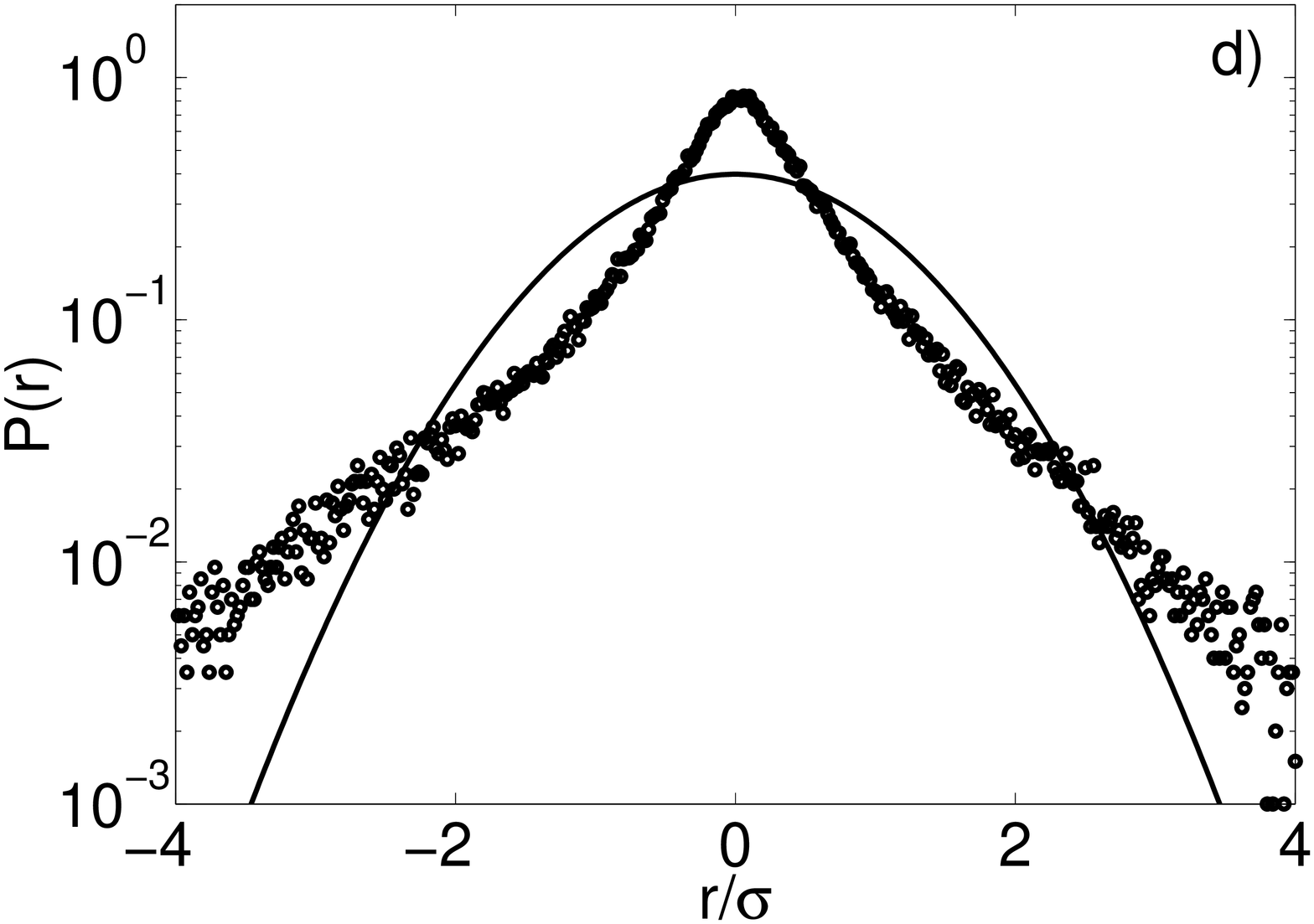} \hfill
\includegraphics[width=5.5cm,angle=270]{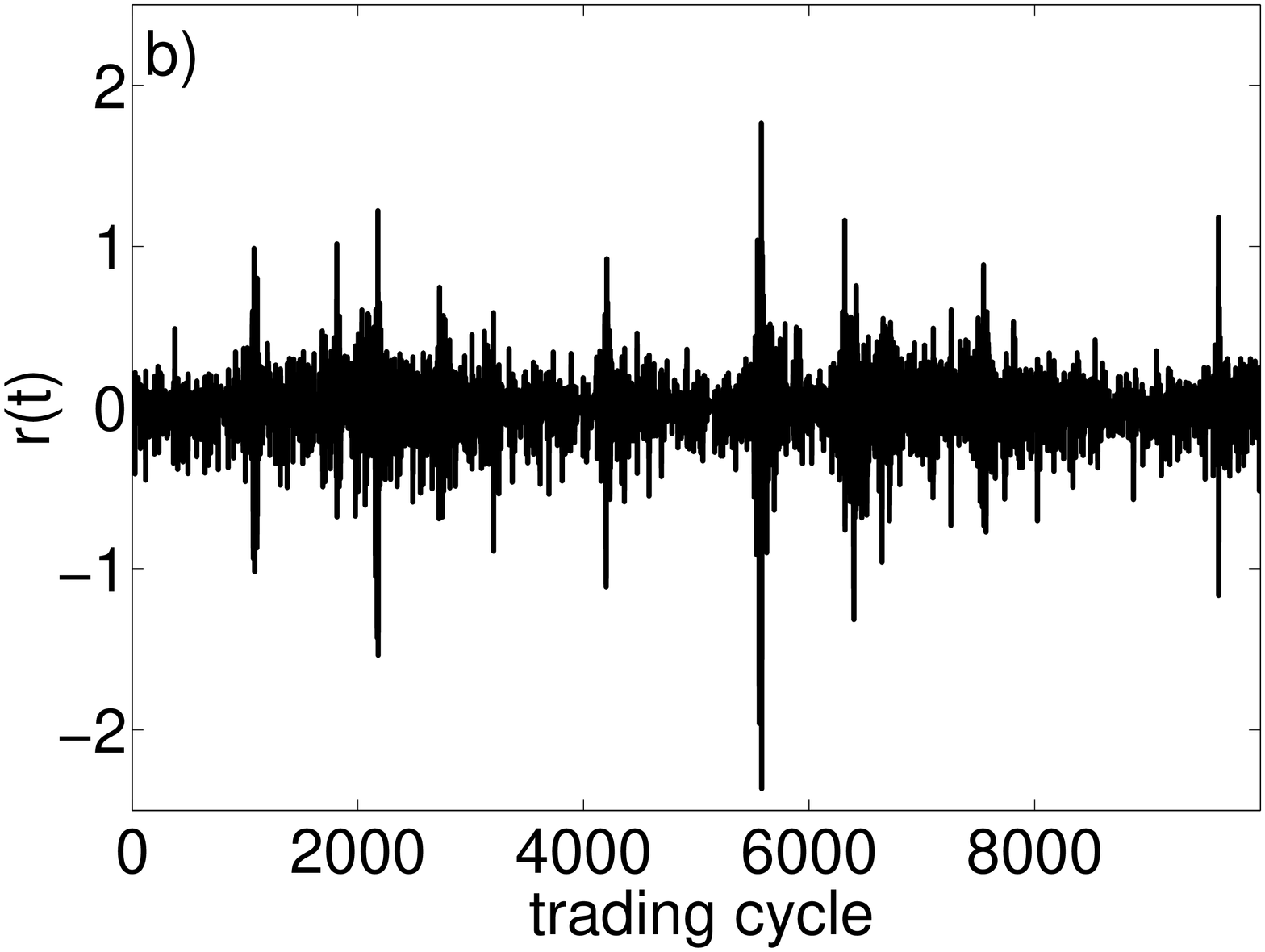} \hfill
\includegraphics[width=5.5cm,angle=270]{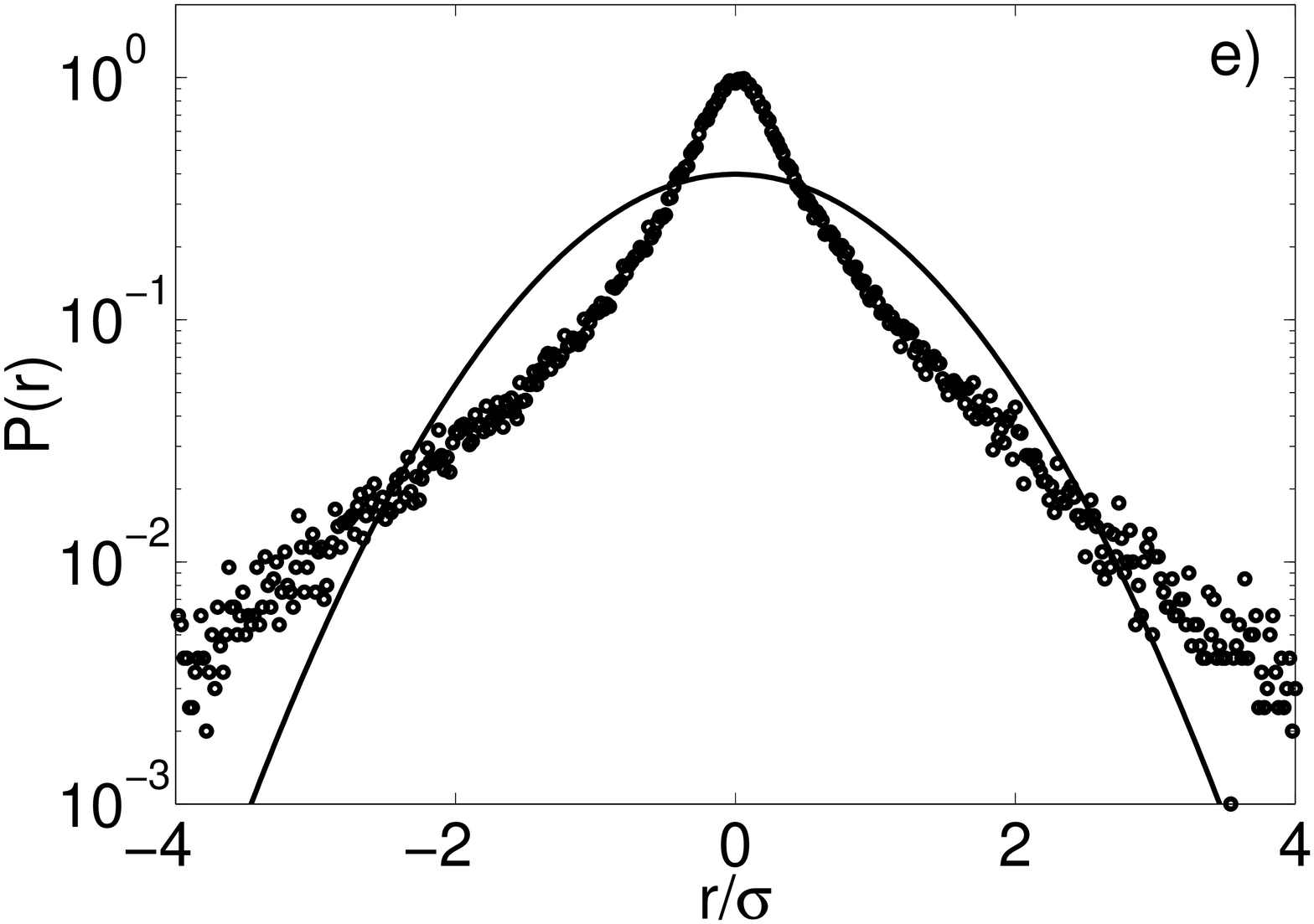}\hfill
\includegraphics[width=5.5cm,angle=270]{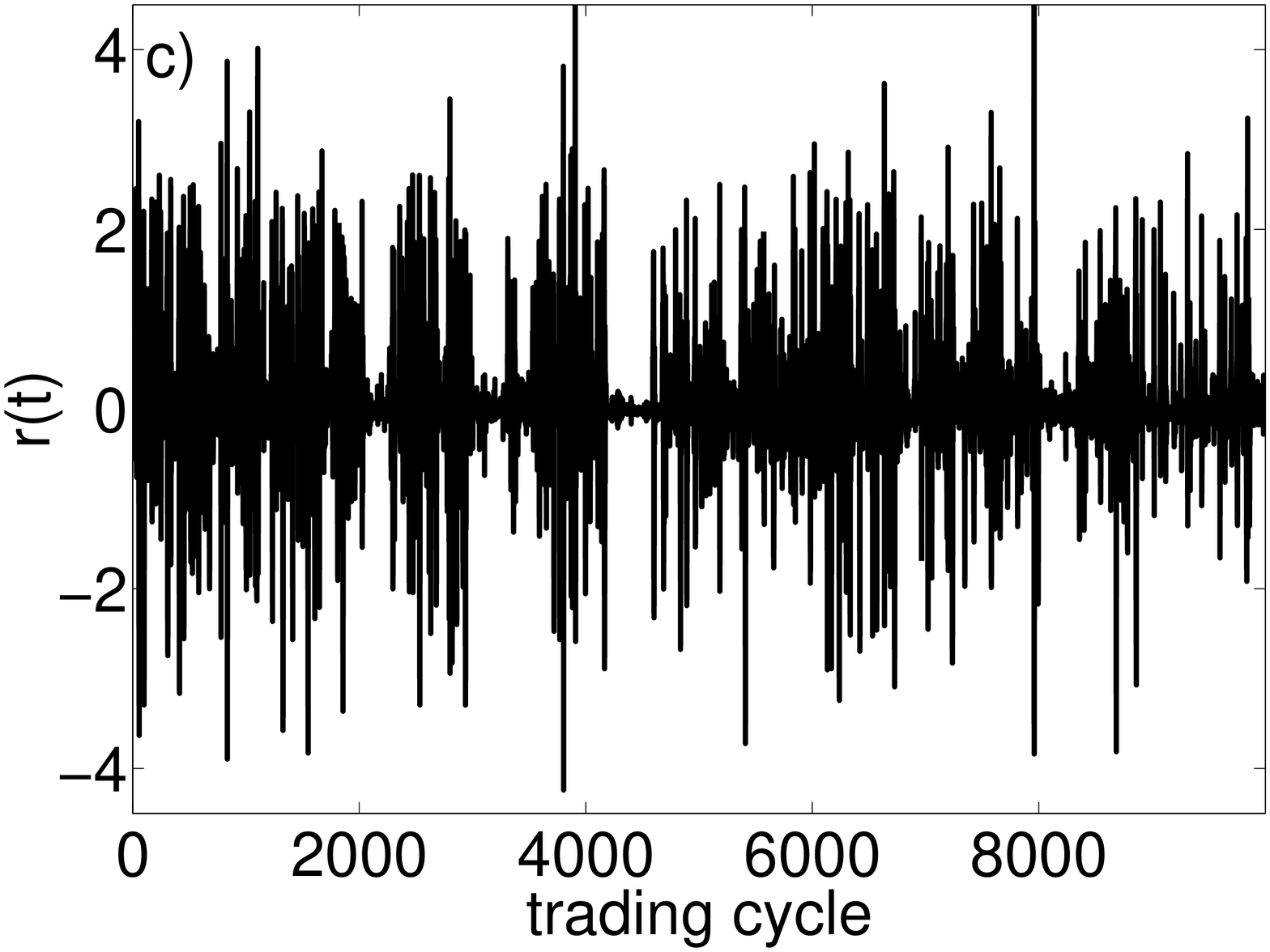} \hfill
\includegraphics[width=5.5cm,angle=270]{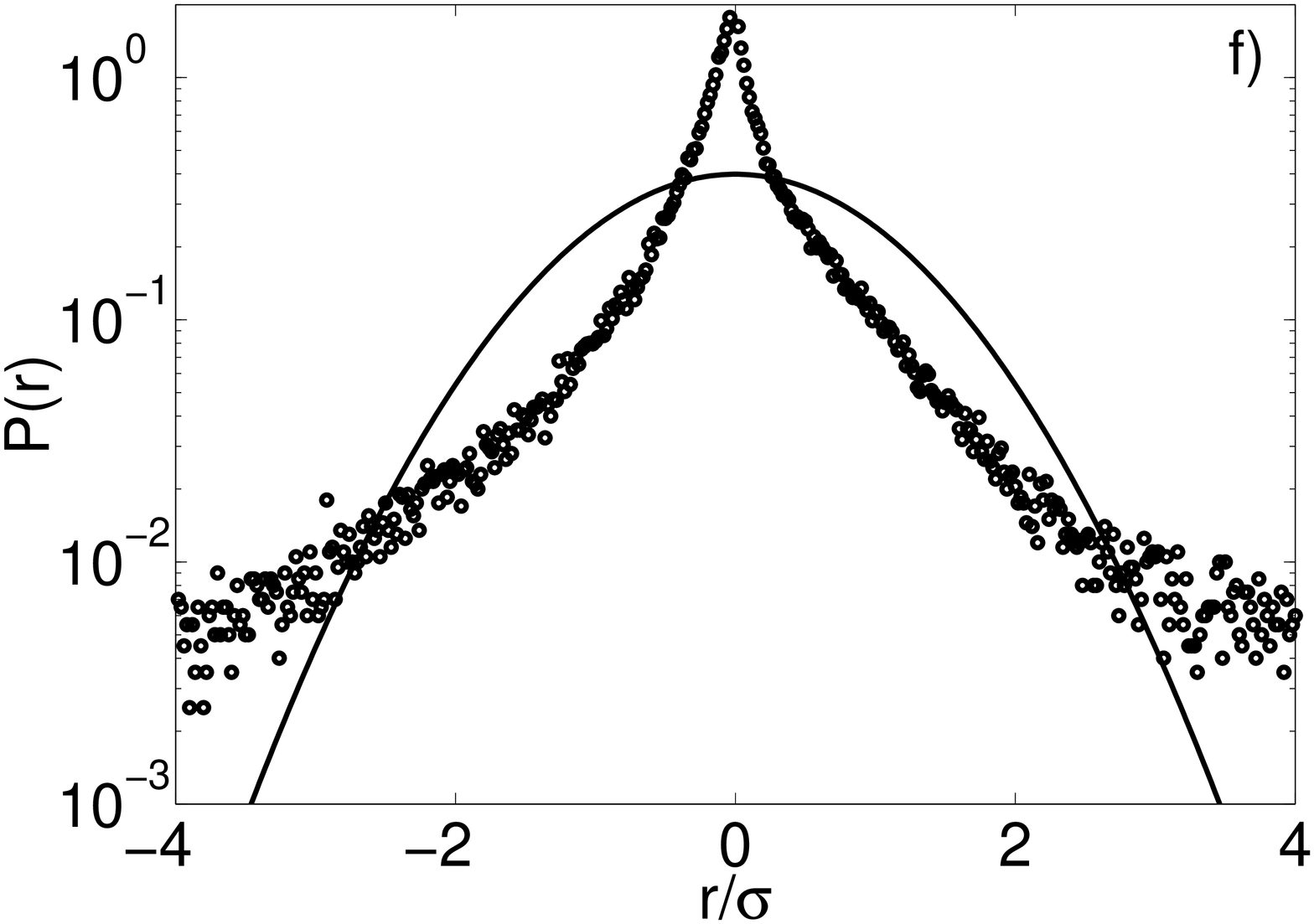} \hfill
\caption{\small Simulation results obtained from different evolutionary models 
 with parameters $N=1000, P=3, S(0)=1000,M(0)=100000, \sigma=1$. After each trading 
 cycle the poorest(upper), a random(middel) or  the richest(lower) agent 
is replaced by a new random one.
Left and right plots are the same as in Fig.\ref{Abb:Basic01} b) and d), respectivly.}
\label{Abb:Evolution}
%\end{center}
\end{figure}

Figs.\ref{Abb:Evolution}a,d show the result for the model in which the
poorest trader is replaced, Figs.\ref{Abb:Evolution}b,e for the model in
which a random agent is chosen, and Figs.\ref{Abb:Evolution}c,f depicts
the effects of each time selecting the richest agent. Surprisingly,
all these evolutionary versions of our model lead to non-Gaussian
fat-tailed return distributions. In particular the tails in
Fig.\ref{Abb:Evolution}f are larger than in
Figs.\ref{Abb:Evolution}b,d.\\
Comparing the correlations of these processes, we see that a
selection of the poorest agent kills trends in the time series
faster and leads to a smaller autocorrelation than in the case with
selecting the richest which exhibits quite strong
anti-correlations.
In order to check the robustness of our results we investigated the
effect of the total number of agents on the results. As discussed
above larger markets employing an evolutionary principle imply a
smaller relative noise level at each cycle. 
We find that also with increasing number of agents the return 
distributions show a strongly non-Gaussian shape. 

\section{Summary and Discussion}

We presented a simple paradigmatic model for a stock market by 
combining game theoretical approaches with a realistic mechanism 
for price determination that reproduces the basic features of returns 
in real price time series. It consists of traders which compete 
by making predictions of future returns. In order to be consistent 
with real trading we found that a two step prediction is necessary 
which we modeled by iterating a linear predictor for each trader. 
To determine the parameters for prediction we used two different 
approaches: 
In our basic model we chose our parameters from a Gaussian 
distribution and change it every time step by adding some noise 
to the parameters. In the evolutionary versions we used different 
kinds of evolution to choose agents and replaced them by new 
random ones.

\begin{figure}[ht]
\begin{center}
\includegraphics[width=9cm,angle=270]{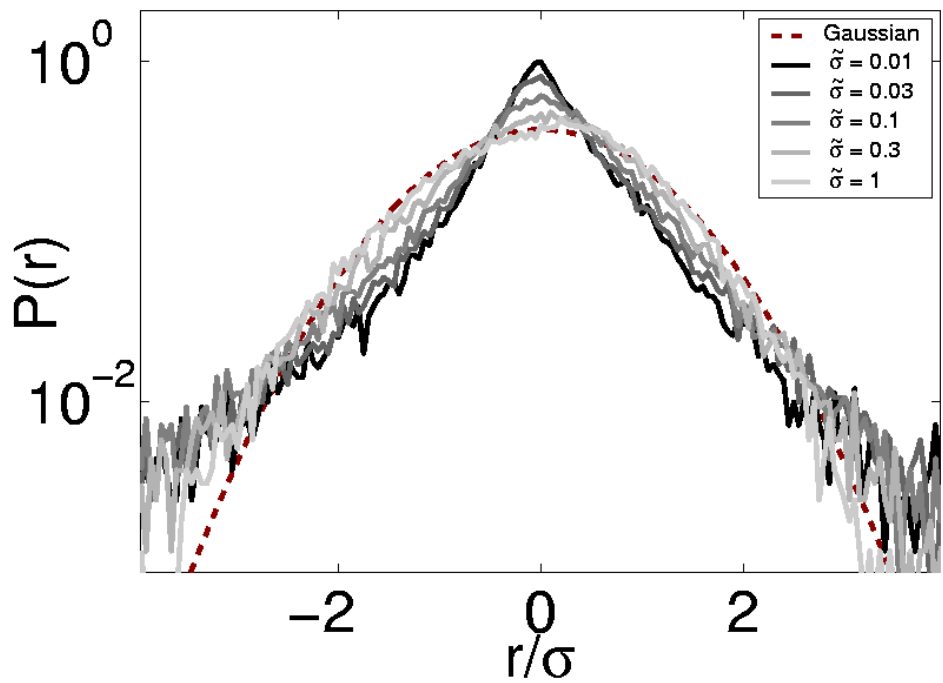}\hfill
\caption{\small Normalized return distribution calculated as in 
Fig \ref{Abb:Basic01} for different values of $\tilde\sigma$ ranging from 0.01 to 1. }
\label{Abb:Multisigma}
\end{center}
\end{figure}

Both models qualitatively reproduce important statistical features of 
returns in real price time series like volatility clustering and 
fat-tails using only a few parameters: the number of agents $N$, 
the initial amount of stocks and cash, the complexity $P$ 
characterizing the agents, and the mean and the standard deviation 
of the parameter distribution. 
In the basic model, the level of an intrinsic Gaussian noise has 
to be specified additionally. As long as this parameter is not to 
small the scaling appears to be realistic (see Fig.\ref{Abb:Multisigma}), 
while below a critical level, the scaling behavior broke down. However, 
our results indicate that in the limit of  $N \to \infty$ the amount 
of noise, necessary for realistic scaling, vanishes. When $\tilde \sigma$ 
is increased the distribution of returns changes via a power law to a Gaussian 
distribution. This shows that the volatility clusters are not caused 
by the mechanism determining the price, but by the memory contained 
in the parameters and, most notably, in the capital of the agents.
In contrast, the noise induced by the evolutionary mechanisms needs no 
adjustment. In some cases the scaling of the return distribution 
became even more algebraic when we decreased the effective noise level 
by increasing the number of agents. The evolutionary models enable 
to study how the dynamics of the agents in parameter space influence 
the statistics of the time series. To this end we modeled three different 
kinds of evolution.  Our study demonstrates that 
strong evolutionary pressure to perform well in a market leads to a destruction
of correlations in the time series, while random selection and, in particular,
anti-evolution induce larger (anti-)correlations. Remarkably, we found 
particularly clear power laws for anti-evolution.

In all variants of our model the results did not critically depend on the
number of agents $N$, and for large $N$ the scalings rather improved. To our
knowledge, our approach therefore represents the first solution to a long
standing problem in modeling markets \cite{Samanidou_02}. Therefore, our model 
could be considered as a prototype of a self organized system, which tends to 
evolve toward a critical state, as one would also expect from real stock markets.
In particular, we think that the combination of a price mechanism with two 
consecutive return predictions and an evolution mechanism that works nearly 
without stochasticity will lead to a deeper understanding of the underlying 
dynamics of stock markets. 

\ack We acknowledge C. van Vreeswijk, M. Bethge and F. Emmert-Streib
for fruitful discussions.

%\bibliographystyle{h-elsevier}
%\bibliography{}

\bibliographystyle{hunsrt}
\bibliography{finance}

%\pagebreak 

%%%%%% Figures 1 %%%%%%%%

%%%%%% Figures 2 %%%%%%%%

%%%%%% Figures 3 %%%%%%%%

%%%%%% Figures 4 %%%%%%%%

%%%%%% Figures 5 %%%%%%%%

\end{document}